\documentclass[11pt]{article}

\usepackage[utf8]{inputenc}
\usepackage[T1]{fontenc}

\usepackage{amsmath,amssymb,amsthm,bbm}

\usepackage{graphicx,transparent,eso-pic}

\usepackage[document]{ragged2e}
\usepackage{pagecolor,color}
\usepackage{soul}

\usepackage{ifthen}

\usepackage{natbib}

\newcommand{\noun}[1]{\textsc{#1}}

\usepackage{url}

\usepackage[margin=2cm]{geometry}

\usepackage{soul}
\soulregister\cite7
\soulregister\citep7
\soulregister\ref7

\usepackage[final]{changes}

\setaddedmarkup{\textcolor{black}{\hl{#1}}}
\setdeletedmarkup{\textcolor{red}{\sout{#1}}}

\usepackage{CJKutf8}
\newcommand{\cn}[1]{
  \begin{CJK*}{UTF8}{gbsn}
  #1
  \end{CJK*}
}

\begin{document}

\title{Evolving accessibility landscapes: mutations of transportation networks in China}
\author{\noun{Juste Raimbault}$^{1,2}$\\
$^1$ UPS CNRS 3611 ISC-PIF\\
$^2$ UMR CNRS 8504 G{\'e}ographie-cit{\'e}s
}
\date{}

\maketitle

\justify

\begin{abstract}
Recent years have witnessed an exceptional extension of public transportation networks in People's Republic of China, both at a national scale with the construction of the first HSR railway network of the world, and at local scales with numerous cities developing high coverage subway networks often from scratch. This chapter studies these mutations, both from a quantitative perspective with the study of the evolution of population accessibility landscapes, at a national level and for several cities, and from a qualitative perspective with fieldwork observations. We confirm that rebalancing planning objectives are well achieved in terms of accessibility at both scales, when all planned lines will be achieved but already in a significant manner. We finally hypothesize possible paths for the coupled network-territory systems, given the extent without precedent of such mutations.
\end{abstract}

\textbf{Keywords : }\textit{Transportation networks; Accessibility; HSR; Urban Transit; Pearl River Delta}

\section{Introduction}

\subsection{Accessibility in territorial systems}

Cities and their future form of large urban metropolitan regions are a well-known gordian knot of the sustainable transitions, as they concentrate many apparent paradoxes of concurrent positive and negative externalities, but also more intrinsically as they are the incubator of social change and innovation \citep{pumain2009innovation} implying a complex role of their physical structure in processes of decoupling growth with environmental and social negative impacts \citep{bergeaud2018bel}. A crucial aspect of the structure of these territorial systems are transportation networks, as activities and flows strongly depend on these at multiple scales \citep{raimbault2018caracterisation}. For example, the development of efficient public transportation systems is crucial to reduce car use, decrease local pollution and global emissions, and enhance geographical equity \citep{sinha2003sustainability}. Accessibility, in its broader sense including effective but also potential access to different amenities \citep{bavoux2005geographie}, is a way to quantify the potential impacts of a transportation network on a territory. The evolution of accessibility landscapes following an evolution of a transportation network is thus a relevant entry into these issues.

China, which is concerned by severe air pollution issues and is a significant contributor to global emissions, has in the last decades initiated profound infrastructural changes to develop inter-city rail transport and urban rail transit, expected to have a significant impact for greenhouse gases emission mitigation \citep{han2008system}, although much remains to be done in terms of total efficiency of the transportation system \citep{chang2013environmental}. \cite{jiao2014impacts} show the role of the High Speed Rail network in the national accessibility, with a remarkable uniformization of accessibility values for a large majority of the population. This can be compared to the evolution of the Chinese railway network during the 20th century as studied by \cite{wang2009spatiotemporal}, for which accessibility was more unevenly distributed. \cite{hou2011transport} establish accessibility changes in the region of Pearl River Delta for different transportation modes. \cite{lyu2016developing} tackle the issue of accessibility in Beijing by classifying territories in the perspective of a Transit Oriented Development, i.e. a territorial development aimed at encouraging the use of public transport.

We study in this chapter the recent transformation of Chinese transportation networks from different point of views, in order to give a broad overview of this mutation processes and their implication in terms of accessibility. The rest of this chapter is organized as follows: we first give some elements of context of the urban environment in which the transportation networks are inserted and the related governance issues. We then develop some case studies, namely road accessibility in Pearl River Delta in relation with the new Hong-Kong-Zhuhai-Macao bridge, urban rail transit accessibility in major cities, and High Speed Rail accessibility at the national level. The next section is aimed at complementing this overview with fieldwork observations. We finally discuss the implication of these results for future trajectories of these territorial systems.

\subsection{Transportation governance and Mega-city regions}

The development of transportation infrastructures is closely related to the governance context in which it is decided. \cite{tang2008impact} show for example the importance of maintaining long-term transport policies in the case of Hong-Kong, to avoid negative externalities as a consequence of a short-term planning. The question of the scale at which planning is done and of governance structure adapted to the actual urban functional and morphological structure is crucial. In particular, a new form of urban development which is interpreted by \cite{lenechet2017peupler} as the most recent transition of settlement systems, namely the emergence of large polycentric urban regions, is closely related to the question of developing integrated territories through infrastructure networks and the governance of these.

The notion of megalopolis has been introduced by~\cite{gottmann1961megalopolis} to describe the emergence of urban agglomerates at a scale that did not exist before. It is at the origin of the concept of \emph{Mega-city Region} (MCR) which was defined by~\cite{hall2006polycentric}. For the European case, they unveil a number of urban regions which are strongly connected regarding mobility flows, connections between companies, which correspond to what they call polycentric \emph{Mega-city Regions} (for example Randstad in Netherlands, the Rhin-Rhur region in Germany). Their characteristics are a certain geographical proximity of centers, a strong integration through flows, and a certain level of polycentrism. It consists in an urban form that did not exist before, which emergence seems linked to globalization processes. This concept is even more relevant with the recent emergence of new types of urbanization, in particular through the accelerated urbanization in countries with a strong economic growth and undergoing a very rapid mutation such as China~\citep{swerts2015megacities}.

\cite{Ye2014200} analyzes the actions of metropolitan governance at the scale of centers of the Pearl River Delta MCR, and more particularly how municipalities of Guangzhou and Foshan have progressively increased their cooperation to form an integrated metropolitan area, what can thus strongly influence the development of transportation for example and allowing the construction of a connected network. A strong tension between bottom-up processes, and a state control which is relatively strong in China, which originates from the Central State, to the province government and local government, has allowed the emergence of such a structure. The competition with other cities in the MCR remains strong, and the logic of integration (in the sense of articulation between the different centers, of interactions and of flows between these) of the MCR is only partly guided by the region. The particular nature of SEZ of Shenzhen and Zhuhai, linked to the privileged relations with the Special Administrative Zones of Hong-Kong and Macao, which have been integrated back to the PRC only at the end of the last millenium and keep a certain level of independence in terms of governance, complicates even more the relations between actors within the region. The issue of a correspondence between some levels of governance and urban processes is a tricky one: \cite{liao2017ouverture} interpret the progressive transfers of economic initiatives from the central power to local authorities as a form of a multi-level governance.

The complexity of these contexts is implicitly crucial in the analyses we will develop in the following: a given network development is generally the outcome of several processes at different scales. A positive accessibility impact at a given scale can have different roles at other scales, and a change in accessibility landscape can be provisory detrimental to ensure a longer-term overall improvement. Therefore, the quantitative analysis of accessibility landscapes we develop in the next section will make more sense when put in context of the fieldwork observations described in the last section.

\section{Evolution of accessibility landscapes}

\subsection{Data and methods}

We propose a simple but multi-scalar and multi-modal illustration of the evolution of accessibility landscapes following major infrastructural change in China. To have a generic measure we consider a temporal potential relative accessibility to populations, defined for a set of geographical objects (that can be for example cities or zones) with populations $P_i$, given a travel time distance matrix $(t_{ij})$ between these areas, and a decay parameter $t_0$ capturing the typical time span of travels, by

\[
Z_i = \frac{1}{\sum_j P_j} \cdot \sum_j P_j \cdot \exp{\left(- \frac{t_{ij}}{t_0}\right)}
\]

This object-level measure of accessibility to other populations can then be summarized for example using an average, but also hierarchy measure such as a rank-size exponent, expressing a level of inequality between areas in terms of accessibility.

We use only populations as territorial data, to ensure a certain genericity at different scales and on different geographical area. At the mesoscopic scale, we use the gridded population dataset at a 1km resolution in 2010 provided by \cite{fu2014grid}. At the macroscopic scale, we use the ChinaCities database, introduced by \cite{swerts2013systemes} and provided by \cite{swerts2017database}. This database has the advantage of ensuring ontologically consistent definitions of cities (in the sense of variable city boundaries capturing evolving urban entities, in this case in the sense of a morphological continuity). Transportation network data are detailed for each case study below. Analyses are done in R, which is relatively flexible to handle GIS data, using a dedicated package for transportation network and accessibility analyses \citep{raimbault2018trpackage}. All source code and data (excluding population data which are openly available from original sources) are openly available on the repository of the project at \url{https://github.com/JusteRaimbault/ChinaAccessibility}.

\subsection{Road accessibility in Pearl River Delta}

\subsubsection{Context of Pearl River Delta}

The first analysis we present enters the specific geographical context of Pearl River Delta (PRD), which is one of the classical illustrations of the structure of a strongly polycentric MCR. We first describe the historical and geographical context of this region. Historically initially only composed by Guangzhou, the development of Hong-Kong and the establishment of Special Economic Zones (SEZ) in the context of opening policies by \noun{Deng Xiaoping}, lead to an extremely rapid development of Shenzhen, and in a less proportion of Zhuhai. Shenzhen and Zhuhai were among the first Special Economic Zones, created in 1979 to attract foreign investments in these areas with flexible economic rules. The development model of Zhuhai was different of Shenzhen, since heavy industry was forbidden. Guangdong province in which PRD is fully located has currently the highest regional GDP within China, and the MCR contains a population of around 60 millions (estimations strongly fluctuate depending on the definition of the MCR which is taken, and the inclusion of the floating population). The phenomenon of migrations from rural areas is highly present in the region and a city such as Dongguan has for example based its economy on factories employing these migrant workers.

\begin{figure}
	\includegraphics[width=0.49\linewidth]{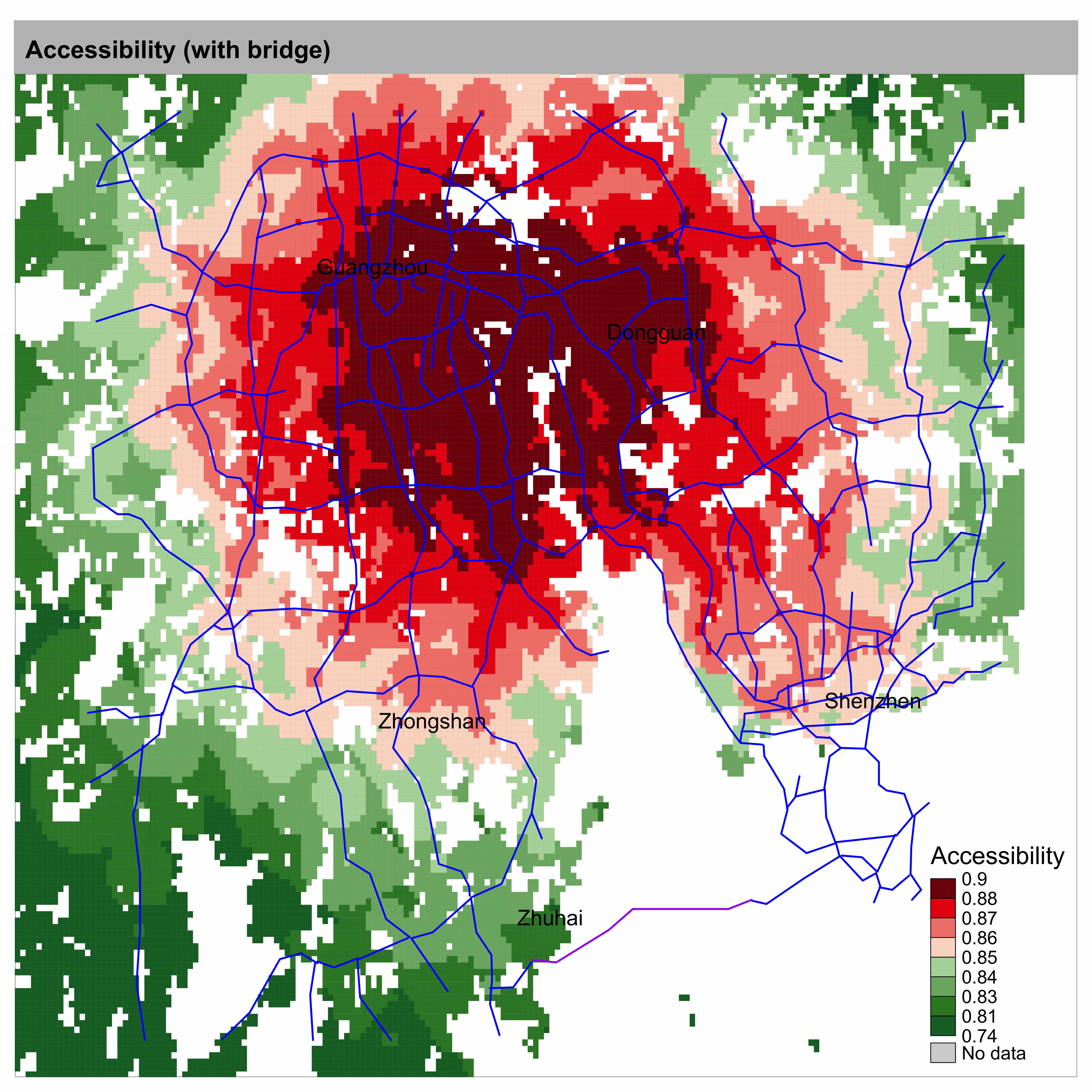}
	\includegraphics[width=0.49\linewidth]{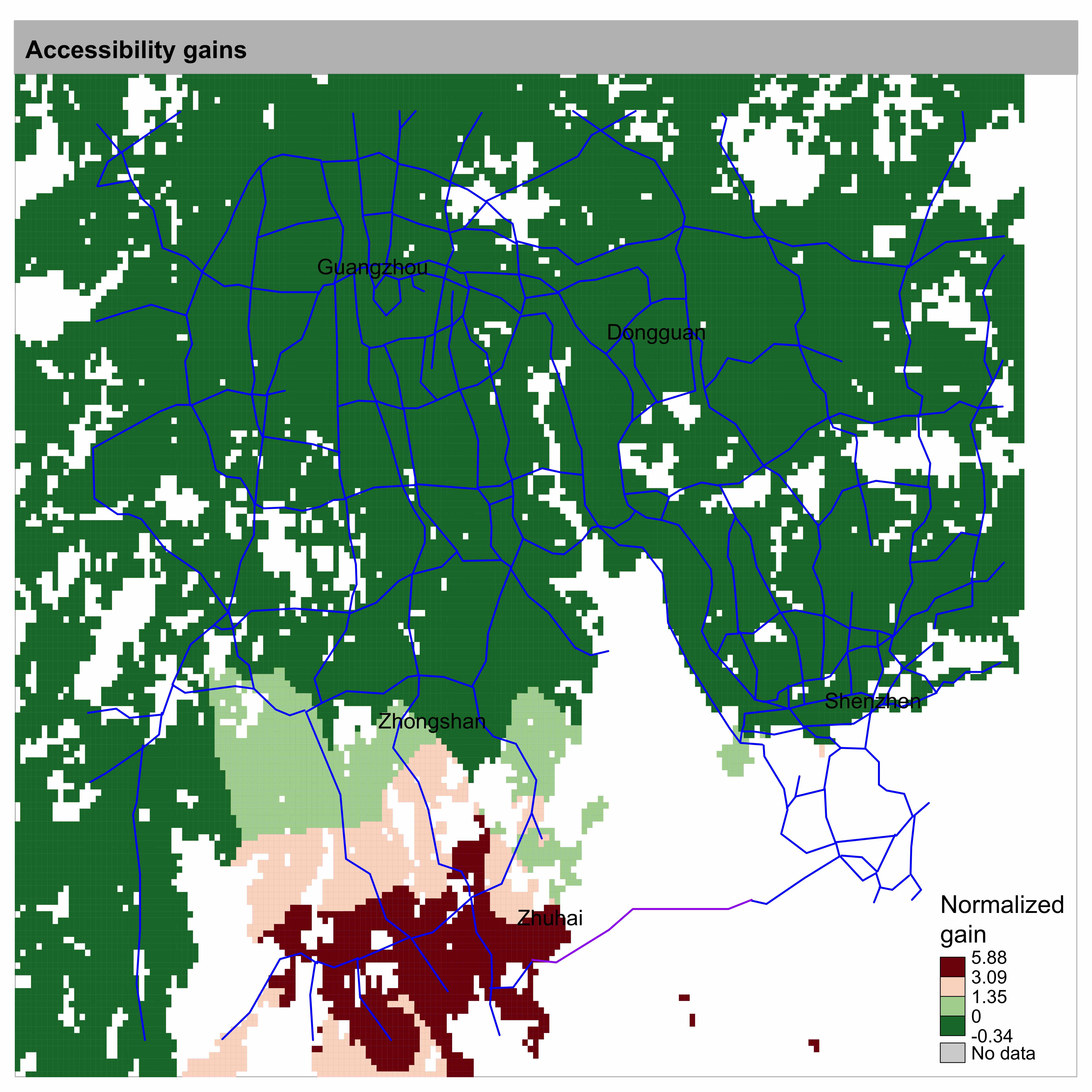}
	\caption{\textbf{Accessibility gain induced by the HZMB in Pearl River Delta, for the territory of mainland China.} (\textit{Left}) Accessibility to population $Z_i$; (\textit{Right}) Normalized accessibility gains. The population of Hong-Kong is taken into account in destination points although it is not included in origin patches. The highway network (2017) is mapped in blue and and the new link of the bridge in purple.\label{fig:casestudies:prd}}
\end{figure}

In the frame of transportation within the Pearl River Delta, there is no specific authority at this scale for the organization of transportation (but indeed entities at the level of the State, of the province and of municipalities), and each municipality manages independently the local network, whereas the connections between cities are ensured by the national train network. This leads to particular situations in which some areas have a very low accessibility, with a very strong heterogeneity locally. Therefore, the southern part of the city of Guangzhou which constitutes a direct access to the sea, is geographically closer to the center of Zhongshan, but a direct link by public transport is difficult to imagine, whereas the area is well linked to the center of Guangzhou by the metro line. A similar situation can be observed at the terminus of line 11 in Shenzhen, for the neighbor district of Dongguan, the latest having a very low accessibility by public transport (see the map in Fig.~\ref{fig:casestudies:prd} for locations, the map giving also the accessibility with the road network). This situation is however transitory, given the infrastructures already being built and the ones planned on a longer term: the Shenzhen metro, which covers today 285km, is planned to reach 30 lines and a length of around 1100km. This implies for Shenzhen a very high future transport density, corresponding to high urban density areas, such that the plan anticipates 70\% of commuting by metro at the horizon 2030, as declared by the official plan of the city~\citep{shenzhen2016plan}. It is clear that these developments mostly follow an existing urban development, a crucial issue is the voluntarism and the capacity to contain urban sprawl and to structure future developments around this new network, in the spirit of a voluntary integration between urbanism and transportation of the type Transit Oriented Development. Different final stations will be connected to the Dongguan metro, and new intercity lines will structure the longer range mobility, what will make the Delta a relatively well integrated in terms of public transport in a rather short temporal horizon. To have an idea of the development of the network in the coming years, the Table~\ref{tab:casestudies:stats} gives the size of the planed networks in the different cities for 2030.


\begin{table}
\caption{\textbf{Public transportation in Pearl River Delta.} We give populations in 2010 taken from \citep{yearbook2013guangdong}. Network lengths are taken from the different planning documents for the Guangzhou metro~\citep{guangzhou2016metro}, the Shenzhen metro~\citep{shenzhen2016plan} and the Dongguan metro~\citep{dongguan2017ditie}, and for the Zhuhai tramway~\citep{zhuhai2016tram}. Zhongshan is not included since it exploits a BRT system but no heavy infrastructure.\label{tab:casestudies:stats}}
\begin{center}
\begin{tabular}{|c|c|c|c|}\hline
City & Population & Network 2016 & Network 2030 \\\hline
	Guangzhou - Foshan & 18.9 Mio & 390km & 800km \\\hline
	Shenzhen & 10.4Mio & 286km & 1124km \\\hline
	Dongguan & 8.2Mio & 38km & 195km \\\hline
	Zhuhai (Tramway) & 1.6Mio & 10km & 173km \\\hline
\end{tabular}
\end{center}	
\end{table}

\subsubsection{Impact of the Zhuhai-Hong-Kong-Macao bridge}

A major transportation infrastructure project in the region is the bridge-tunnel closing the mouth of the Delta, linking Zhuhai and Macao to Hong-Kong (HZMB). The length of the crossing is 36.5km, what makes it an exceptional infrastructure~\citep{hussain2011hong}. The opening to traffic was delayed by several years and was finally achieved in October 2018 (see the official website at \url{http://www.hzmb.org/cn/default.asp}). \cite{zhou2016medium} shows that the expected changes in accessibility patterns for the West of the Delta are relatively strong, and these can potentially induce strong bifurcations in the trajectories of cities. The necessity of the project is advocated by the different stakeholders of the project (Guangdong province, Hong-Kong Special Administrative Region, Macao Special Administrative Region) using an argumentation of a strong economic benefit in the frame of opening policies, and also through a social benefit for the West in particular. For example, Zhuhai is positioned as a new pivot between Hong-Kong and the West. The balancing of accessibility, in the sense of a diminution of spatial accessibility inequalities, operates however only for the private car transportation mode, what conducts to question its potential impacts: on the one hand the access to automotive remains reserved to a part of the population only, on the other hand the negative impacts of congestion can rapidly moderate the accessibility gains. These accessibility gains are mapped following the method described above, and shown with accessibility $Z_i$ itself in Fig.~\ref{fig:casestudies:prd}. We used here the freeway network within the region in 2017, and added the bridge link. We observe that although population accessibility remains the lowest in Zhuhai compared to Guangzhou and Shenzhen, it is where the relative gains are the highest and the bridge operates a significant spatial rebalancing.

\subsection{Public transportation accessibility in main cities}

\begin{figure}
	\includegraphics[width=\textwidth]{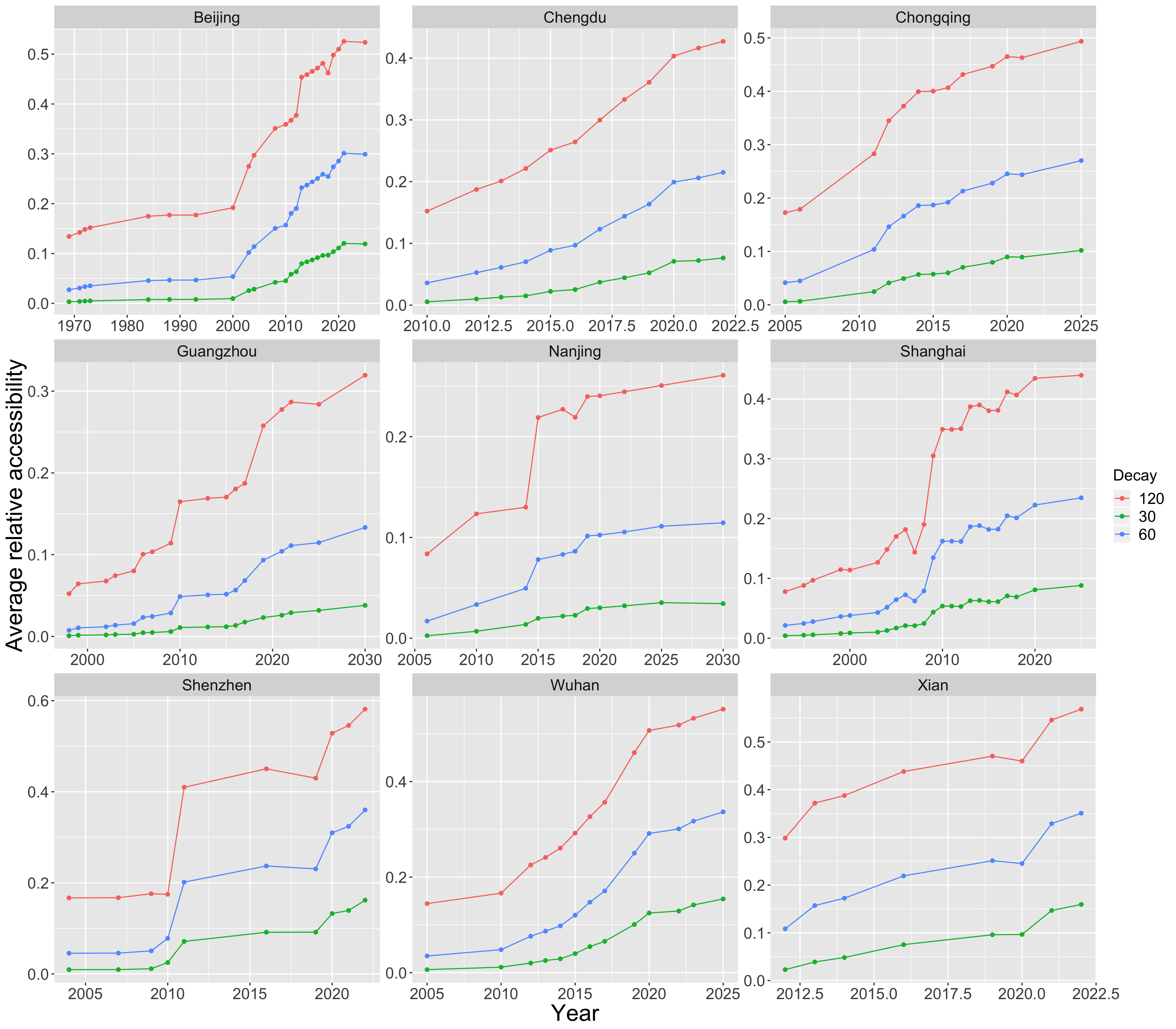}
	\caption{\textbf{Evolution of the average urban rail transit accessibility for main Chinese cities.} For the 9 studied areas, we plot as a function of year the average population accessibility with the 2010 patch-level population, for different values of the time decay parameter.\label{fig:tcaccessavg}}
\end{figure}

\begin{figure}
	\includegraphics[width=\textwidth]{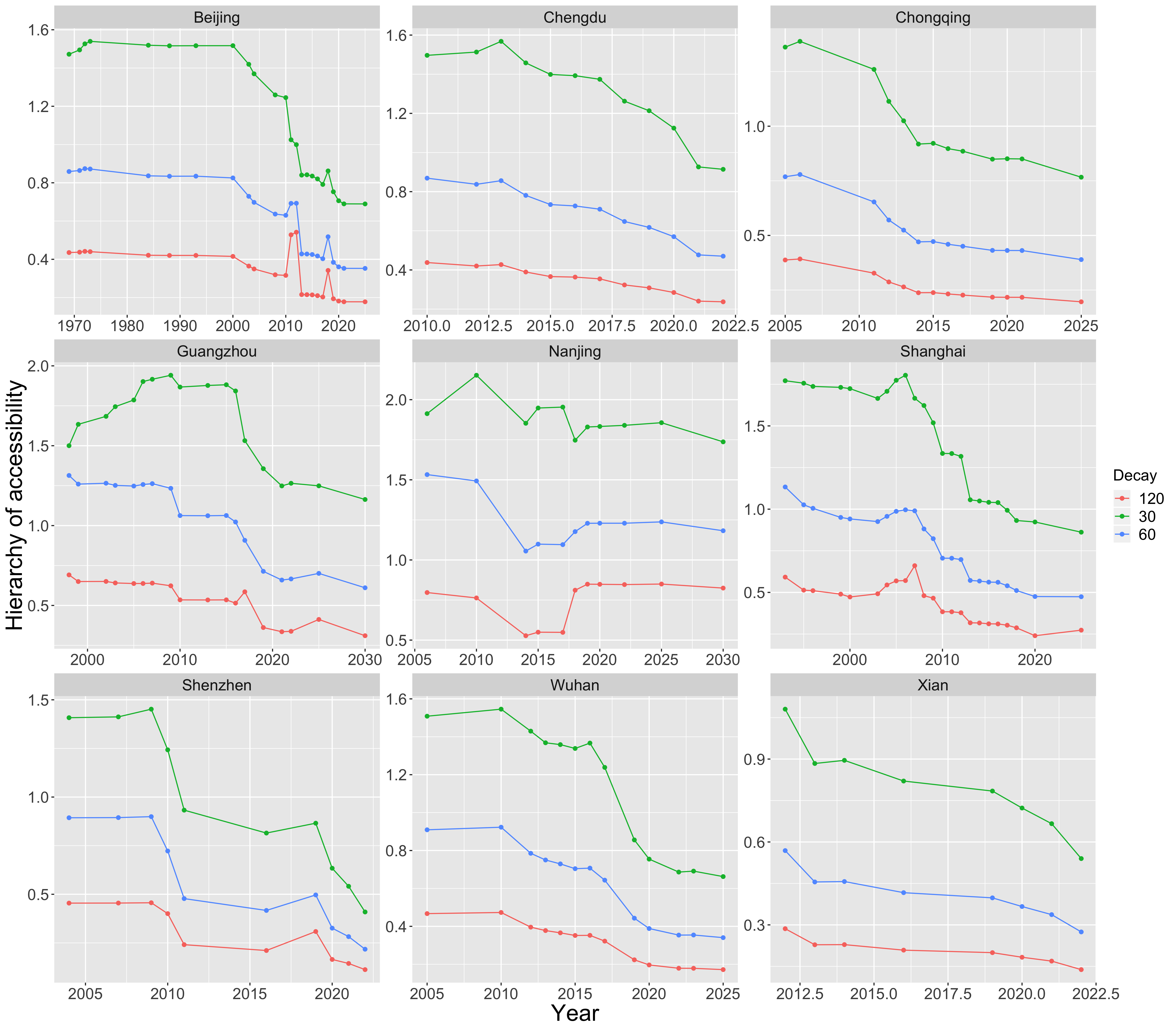}
	\caption{\textbf{Evolution of the hierarchy of urban rail transit accessibility for main Chinese cities.} We plot the rank-size coefficient for accessibilities with the same settings as in Fig.~\ref{fig:tcaccessavg}.\label{fig:tcaccesshierarchy}}
\end{figure}



We now consider the evolution of urban rail transit accessibility in larger cities, considering currently existing networks, lines in constructions and planned lines. As 39 cities have existing, in construction, or approved metro networks, we consider the largest systems in terms of ridership with an annual ridership in 2017 larger than 600Mio passengers. This correspond to 9 cities (Beijing, Shanghai, Nanjing, Guangzhou-Foshan, Shenzhen, Wuhan, Xian, Chengdu, Chongqing) taken in the broader sense of their metropolitan areas. The metro networks are vectorized from city maps (OpenStreetMap and Baidu Ditu) and opening years are taken into account, using history of metro systems from wikipedia Dates up to 2022 correspond to lines already in construction (included in Baidu Ditu), whereas different planning documents are included for lines planned later (to validate or invalidate the wikipedia information which is in some cases outdated; furthermore lines for which no explicit plan in terms of corridor or date is given are not included to avoid too much uncertainty). We consider here population accessibility at the 1km patch level. Only the metro network (average speed of 50km/h) is considered and connectors (speed of 10km/h) are used to connect each patch to the nearest station.


We show in Fig.~\ref{fig:tcaccessavg} values of average relative accessibilities for each city, for different values of the decay parameter $t_0 \in \{30,60,120\}$ minutes (typical travel times in a metropolitan area). It is important to note that most areas have already reached a ``mature'' state, in terms of the potential accessibility gains for the baseline population. As we do not use projected populations and that population growth will most probably occur around transit stations (at least in several districts where TOD policies are implemented), the gains we estimate are underestimated, and can reasonably be interpreted as a lower bound for the relative accessibility. Regarding the evolution of hierarchies, given here by $\alpha$ if $\log Z_i = \beta - \alpha \cdot \log r_i$ with $r_i$ ranks in decreasing order, we observe in Fig.~\ref{fig:tcaccesshierarchy} on the long term a significant decrease in accessibility inequalities, for all cities and all decays. On shorter time spans, some cities such as Guangzhou or Shanghai present a provisory increase in hierarchy, corresponding to situations where the opened line connect more populated areas, but with a large number of areas with a low accessibility. Most cities have a decreasing hierarchy for all decays, at the exception of Nanjing for which the differential are rather low, and for which there is even an increase for the longest travel time, suggesting a different metropolitan structure.

Although this analysis remains limited for several reasons (including the absence of congestion, fixed population, and accessibility to population only), it however shows that these large metropolitan areas have already highly efficient networks and that accessibility inequalities mostly significantly decrease thanks to the development of these metro network.

\subsection{City-level accessibilities with the rise of High Speed Rail}


Finally, we study the population accessibility at a national level considering the construction of the High Speed Rail (HSR) network. Since 2008, the People's Republic of China has established from scratch the largest HSR system in the world, witnessing a strong policy for territorial development, sustainability and territorial equity. With this regard, HSR is an important component of the new doctrine of the State Party (``Socialism with Chinese Characteristics''), recalling the importance of governance structures in the evolution of such infrastructure networks.

We vectorized the HSR network (existing and in construction, from Baidu Ditu and OpenStreetMap) to obtain a dynamic network spanning between 2008 and 2021. We couple it with the ChinaCities database~\citep{swerts2017database}, considering cities of more than 500,000 inhabitants in 2010. We use only one date if the database as the latest available are 2000 and 2010. To obtain reasonable accessibility landscape, we abstract the traditional railway network with a full network between all considered cities with a speed of 100km/h, and overlay the growing HSR network with real speeds (ranging between 200km/h and 350km/h).

\begin{figure}
	\includegraphics[width=0.49\textwidth]{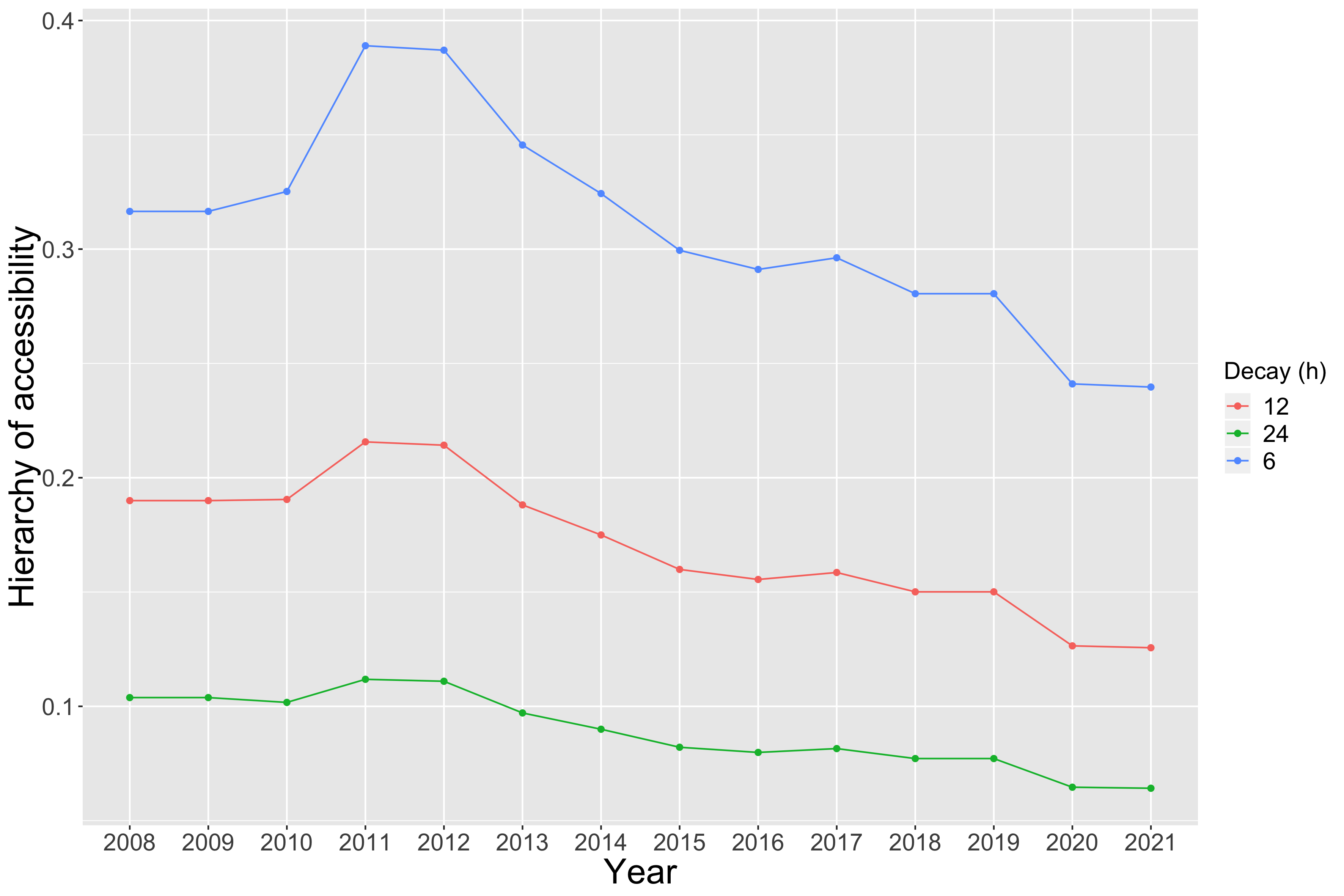}
	\includegraphics[width=0.49\textwidth]{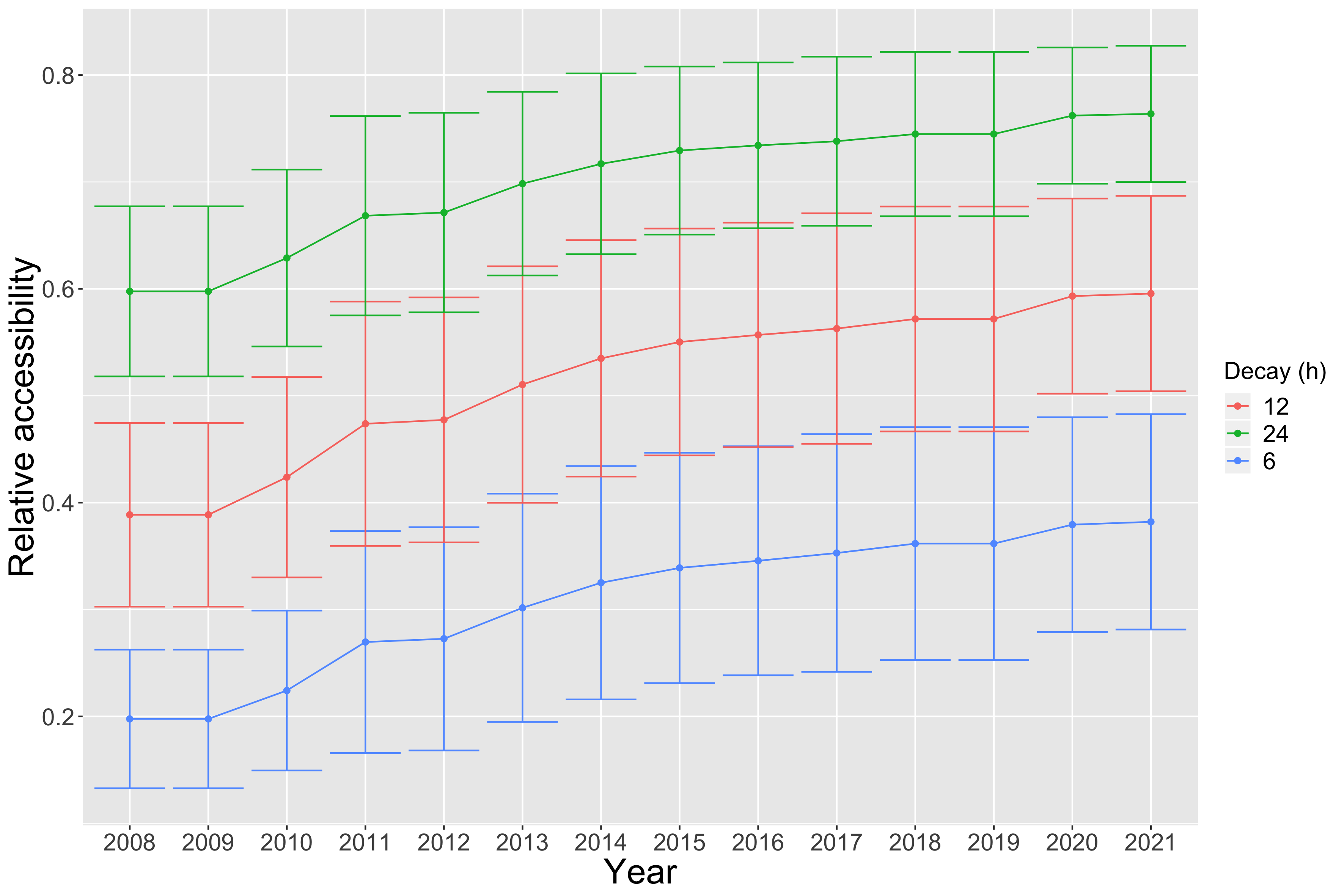}
	\caption{\textbf{Evolution of summary measures of city-level accessibility following the development of the high-speed rail network.} (\textit{Left}) Hierarchy of the distribution of accessibility, given by the rank-size exponent, for different values of the time decay; (\textit{Right}) Corresponding average value of accessibility, with standard deviation of the distribution over cities as error bars.\label{fig:hsraccess}}
\end{figure}

We show the temporal evolution of accessibility measures in Fig.~\ref{fig:hsraccess}, which have been computed for typical travel time at the national scale (6, 12 and 24 hours). We observe for all decays a decreasing trend of the hierarchy over the all period, confirming that accessibility rebalancing is efficiently achieved by the HSR network. However, this decrease is not immediate and a significant peak, especially for the lowest travel time, appears in 2011 and 2012. This correspond to the dates when the Beijing-Shanghai line (2011) and the Beijing-Guangzhou line were fully open. At that time, as these large cities are far, the high decrease in travel time lead to a high increase of accessibility for these larger cities, whereas other territories were left behind. Later, with the maturation of the network, accessibility inequalities were eventually reduced. This dynamic has implications for the choices to be made when developing a new network, showing that the longer-term goal must be kept in mind when judging the investment choices and that middle-term gains for larger cities can be compatible with long-term territorial equity, contradicting the critics on high speed lines connecting large cities and neglecting forgotten territories. Regarding average accessibility gains, the construction of the full network correspond to a significant increase for all decays, and in 2018 most of the gains have already been attained.

We have shown with different case studies, for different transportation modes and at different scales, what are the dynamics of accessibility landscapes associating to the growth of transportation infrastructure networks, and how Chinese planning has achieved accessibility gains and balancing without precedent (given e.g. the length and extent of the HSR, but also the speed of construction of the metro networks) at different scales. This quantitative knowledge can furthermore be illustrated with concrete situations in practice, that we detail in the next section.

\section{Qualitative fieldwork observations}

This section proposes to illustrate the issue of interactions between transportation networks and territories, and more particularly issues related to accessibility. To do that, we show a diversity of possible situations that can be perceived in a qualitative way at the microscopic scale, through concrete fieldwork examples. The geographical subject is mostly Pearl River Delta, in Guangdong province, that we already described before, and more particularly mostly the city of Zhuhai. The objective is to enrich the previous analysis with concrete situations and observations.

We assume the term of \emph{Geographical Fieldwork}, with all knowledge of epistemological debates its use can raise. Indeed, we extract observations from places that were experimented, in the context of a given problematic~\citep{retaille2010terrain}. Our approach will also highlight the role of representations, underlined as a type of fieldwork in itself by~\cite{lefort2012terrain}, when we will give a subjective view.

The fieldwork was achieved in the frame of the European project Medium which established a partnership between European and Chinese universities. The project was entitled ``\textit{New pathways for sustainable urban development in China’s medium-sized cities}''. It aims at studying sustainability through an interdisciplinary and multidimensional viewpoint, in the case of rapidly growing urban areas, by concentrating on Chinese medium-sized cities. Three medium-sized Chinese cities were chosen as a case study. The definition of medium-sized cities considered for the project is broader than the official statistical definition of the Chinese government, and covers cities from 1 to 10 millions of inhabitants. More information is available on the website of the project at \url{http://mediumcities-china.org/}. In this context, Zhuhai was chosen as a case study. When the source is not explicitly given, observations come from fieldwork, for which detailed narrative reports and photographies are available on the open repository of this work at \url{https://github.com/JusteRaimbault/CityNetwork/tree/master/Data/Fieldwork}.

The format of narrative reports is ``on-the-fly'' following the recommendations of \cite{goffman1989fieldwork} for taking notes in an immersive fieldwork in particular, whereas the voluntary subjective position (mostly in the detailed interviews described by \cite{raimbault2018caracterisation}) follows \cite{ball1990self} which recalls the importance of reflexivity in order to draw rigorous conclusions from qualitative fieldwork observations of which the researcher is a part in itself. The consideration of the researcher as a \emph{subject} in relation with its object of study does not imply in our case a feedback of the researcher on the system because of its size in the case of a transportation network at the scale of the city, and indeed a conditioning of observations by a subjectivity of which we must detach in the posterior exploitation of the observation material, but which ignoring can only increase the biases.

\subsection{Development of a transportation network}

The objective of fieldwork is thus to observe the multiple facets and layers of a complex public transport system which is always transforming, its links with observable urban operations, and to what extent these witness of interaction processes between networks and territories. The spatial extent of observations spans on Zhuhai as an illustration of local transportation but also punctually on other regions in China. These observations have their proper logic in comparison to the modeling of transportation networks or data analysis, such as accessibility studies or interaction models between land-use and transportation, that will be done in the following. Indeed, these fail generally in capturing aspects at a large scale, which are often directly linked to the user, and which can become crucial regarding the effective use of the network. For example, multi-modality is profoundly transformed by the use of the network. Multi-modality consists in the combination of different transportation modes: road, train, metropolitan, tramway, bus, peaceful modes, etc., in a mobility pattern. A multimodal transportation system consists in the superposition of modal layers, and these can be more or less well articulated for the production of optimal routes following multiple objectives (cost, time, generalized cost, comfort, etc.) which themselves depend on the user, and of the mobility pattern. It can in practice made efficient through the emergence of self-organized informal transportation modes, or the establishment of new modes such as bike-sharing, what solves the ``last-mile problem''~\citep{liu2012solving}, which seems to be often neglected in the planning of newly developed areas in China. On the contrary, practical details such as tickets reservation or check-in delays at boarding can considerably influence use patterns.


Several trips on the Chinese territory were made to observe the concrete manifestations of the high speed network development. Since its inception in 2008, the Chinese HSR network has a great success and lines are currently saturated. It answers primary demand patterns in terms of city size, showing that it was planned such that the network answers to territorial dynamics. Its high usage shows the impact of network on mobility, what is a possible precursor of territorial mutations.

To show to what extent territories can influence the development of network in diverse ways, we can take a particular example, linked to the development of tourism, which corresponds to a particular dimension taken into account in planning. Thus, the line between Guangzhou and Guiyang (North-West axis which is precursor of the future direct link Guangzhou-Chengdu) have witnessed the opening of stations specifically for the development of tourism, such as Yangshuo in Guangxi, which number of visits has then strongly increased. One year after the opening of the station, the main road link with the city is still under construction, showing that the different networks react differently to constraints at different levels. A higher number of trains stops on week-ends - more than one each hour, are are full more than two weeks in advance. New mobility patterns can be induced by this new offer, as illustrate the interview of an inhabitant of Guangzhou done in Yangshuo, which came for a short week-end with her colleagues, in the context of a ``team-building'' trip financed by her startup in information technology. These new mobility practices are shown in a second interview of an inhabitant of Beijing met at Emeishan, sent by her company in Industrial Design for a short stay in Chengdu for a training in a local subsidiary. The company prefers the high speed train, and it recently increased the mobility practices for its employees.

A similar strategy can be observed concerning the connection of touristic destinations for the line Chengdu-Emeishan. The principal objective of this line is for now to serve the highly frequented touristic destinations of Emeishan and Leshan. However, the missing link between Leshan and Guiyang is already well advanced in its construction and will complete the direct link between Guangzhou and Chengdu. This reveals diachronic and complementary dynamics of network development following properties of territories. This line is a part of the structuring skeleton of the ``8+8'' recently reformulated by the central government, which corresponds to the general plan for future high speed lines, recently actualized to include 8 North-South parallels and 8 East-West others, completing the 4+4 already realized.  The traversed territories expect a lot from it as show \cite{lu2012chengdu} for the city of Yibin halfway between Chengdu and Guiyang.

\begin{figure}
	\includegraphics[width=\linewidth]{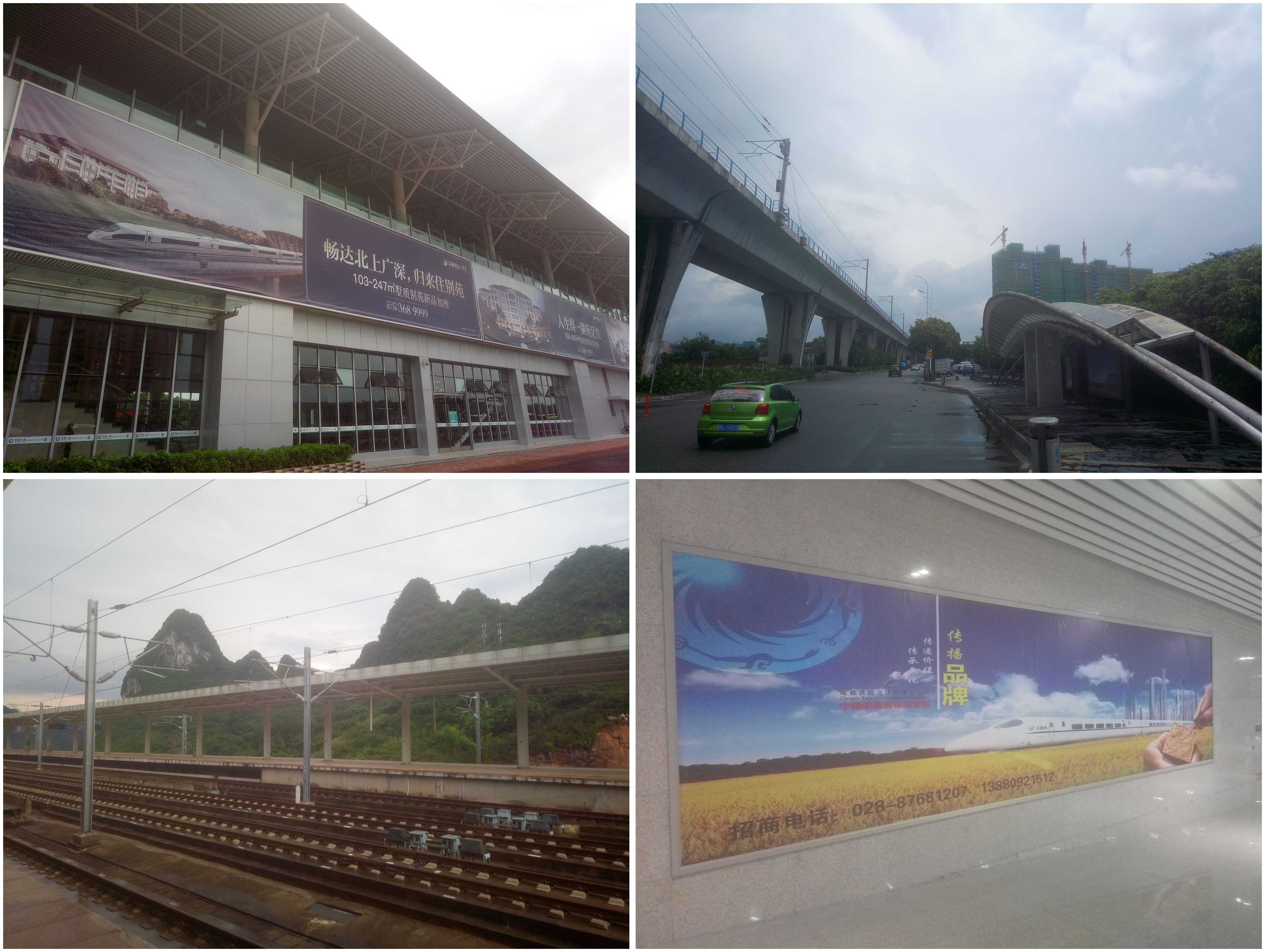}
	\caption{\textbf{Local manifestations of the mutations induced by the new high speed network.} \textit{(Top Left)} High speed station of Tangjia, in Zhuhai city. The monumental advertisement for a real estate operation praises the merits of the proximity to the network, which is also used as an argument for higher prices; \textit{(Top Right)} High speed line in Zhuhai, deserted bus stop and real estate project being realized in a difficultly accessible area: this urban fringe is in direct contact with the rural environment on the other side of the line, and eccentric from the city; \textit{(Bottom Left)} Yangshuo station on the Guangzhou-Guiyang line, which principal function is the development of this touristic destination which bases most of its economy on that field; \textit{(Bottom Right)} Advertisement for high speed in Sichuan, at the station of the international Chengdu airport on the line to Leshan and Emeishan. The train departs from the futurist city to fly over the countryside, recalling the tunnel effect of territories telescoped by high speed.\label{fig:qualitative:hsr}}
\end{figure}

We also observe join mutations of the railway network and of the city. We illustrate thus in Fig.~\ref{fig:qualitative:hsr} the insertion of the HSR in its territories. Direct effects of the network are linked to the development of totally new districts in the neighborhood of new stations, sometimes in an approach of type ``\emph{Transit Oriented Development}'' (TOD, which as we recall is a planning paradigm which aims at articulating the development of an heavy transportation infrastructure with urbanization, typically through a densification around stations). Furthermore, more subtle indirect effects are suggested by clues such as the promotion of operations through advertisement. It shows the socio-economic expectations regarding the network and the local agents which have to contribute to its success: advertisements claiming the merits of high speed, and the selling of appartements in the associated real estate operations. This dynamic seems to contribute to the construction of a ``middle class'' and of the role it has to play in the dynamism of territories~\citep{rocca2008power}, construction which is as much concrete since it depends on objective realities, as imaginary in the academic and political discourse, which construct the object simultaneously to its study or use. The insertion of lines in territories seems in some cases to be forced, as shows the Yangshuo station which exploits the tourism opportunity offered by the passage of the line in a low populated area but which is very attractive by its landscapes, or the new real estate operations in Zhuhai which are not very accessible because of their price.

Finally, it is important to remark that the network development answers simultaneously to different types of territorial contexts. Branches of the new high speed network with a short range, such as the line Guangzhou-Zhuhai, can be seen as being at the intermediary between a long range service and a proximity regional transport, depending on the modularity of serving patterns. This line is thus placed within long range urban interactions (the service Zhuhai-Guiyang being for example ensured) and within interactions in the mega-city region, most of the service being trains to Guangzhou. To this can be added the classical train network which keeps a certain role in territorial interactions: some connections require the use of both networks and of urban transportation, such as the link between Zhuhai and Hong-Kong, experimented through terrestrial transportation modes only. Indeed, following the Hato Typhoon on 23/08/2017, maritime links with the center of Hong-Kong and the international airport has been interrupted for a significant part of the delta, and has been reopened for Zhuhai in the beginning of November 2017 only, inspiring a fieldwork trip using terrestrial public transportation.

\subsection{Implementing TOD: contrasted illustrations}

The simultaneous development of the transportation network and the urban environment can be directly observed on the field. The local urban network and real estate development operations are planned closely with the new train network: the Zhuhai tramway, for which a single line is open at the current time and still being tested, is thought to participate in a TOD approach (see e.g. preliminary works of planning consulting, such as for example \url{https://wenku.baidu.com/view/b1526461ff00bed5b8f31d01.html} for the context of the new Xiaozhen district, in the West of Xiangzhou) to urban development which aims at favoring the use of public transportation and a city with less cars, such as wanted for example by the Planning Committee of the \emph{High-Tech Zone} in charge of the development around Zhuhai North station. The observation of the surroundings of Tangjia station, also built in the same spirit, reveals a certain atmosphere of desertion and an unpractical organisation can lead to questioning the efficacy of the approach. This also suggests a certain self-fulfilling nature of the project, as suggested by advertisements for new real estate for sale, insisting on the importance of the presence of the railway line. A full narrative encouraging local actors and individuals to be involved around TOD seems to be used by different actors of development.

\begin{figure}
	\includegraphics[width=\linewidth]{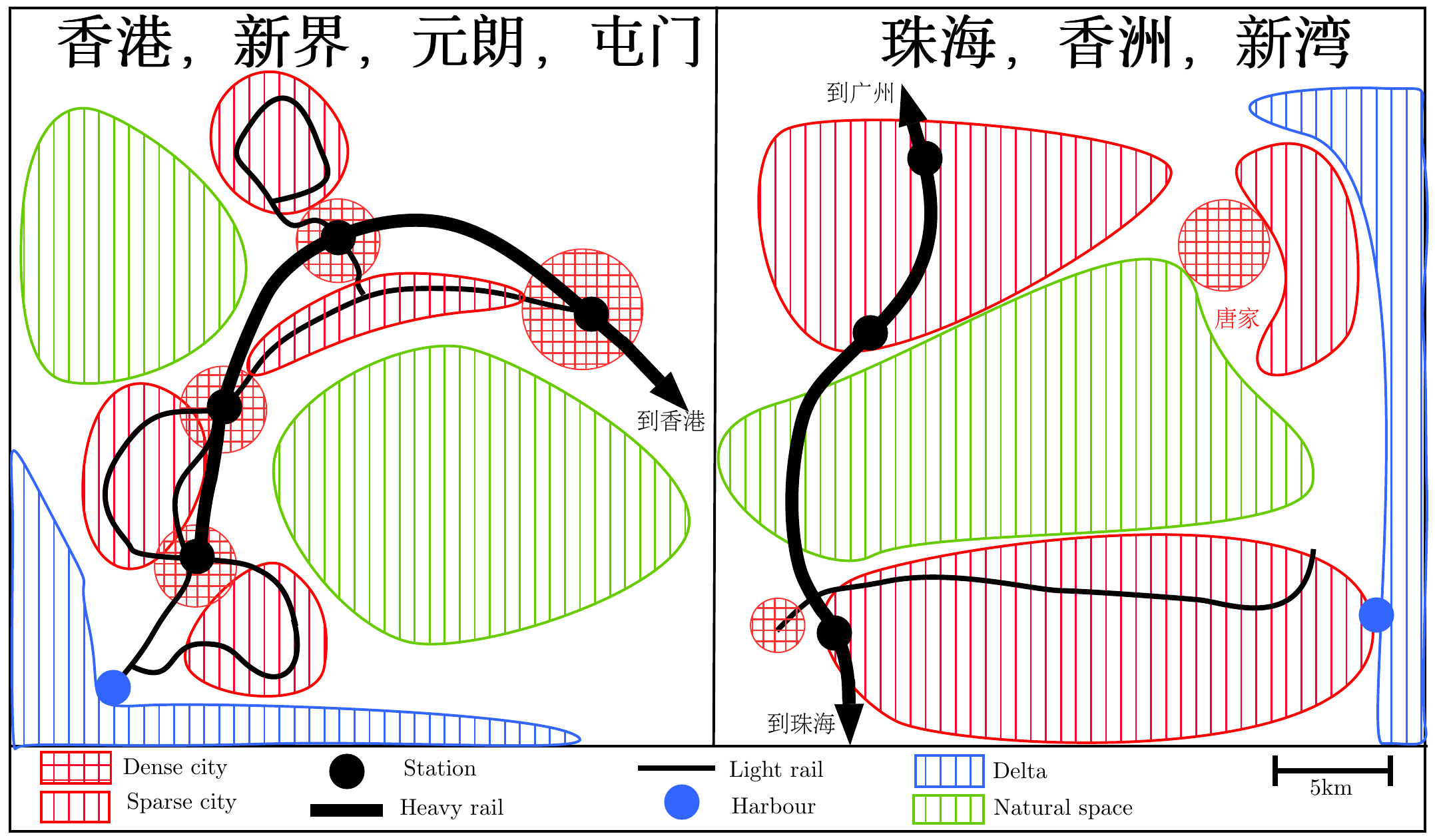}
	\caption{\textbf{Comparative analysis of two implementations of TOD in PRD.} At a comparable scale, we synthesize the urban configuration of Yuenlong and Tuenmun, Hong-Kong New Territories, on the left, and of Xinwan, Xiangzhou, Zhuhai on the right, which contains the Zhuhai High-tech zone in its Northern part in particular. The configurations illustrate different dynamics of articulation, and shifted construction temporalities, unveiling thus different realities under the notion of TOD. A first interpretation would be that it is effective if the trajectory of the full territorial system (urban development and transportation network) is modified early in its genesis, whereas a system with a higher level of maturity will have more inertia.\label{fig:qualitative:schema}}
\end{figure}

Other fieldwork observations, such as in the \emph{New Territories} in Hong-Kong, witness of an efficient TOD which fulfils its objective, with a complementarity between heavy rail and local light tramway, and also a high urban density around stations. These observations recall the complexity of urban trajectories coupled to the development of the network, and that we must remain cautious before drawing any general conclusion from particular cases. We summarize in Fig.~\ref{fig:qualitative:schema} the comparison between the two TOD cases detailed above, as synthetic schemes of urban structures of each area. In Hong-Kong, urban areas have been conjointly planned with the MTR line (heavy transport) and the multiple light tramway lines~\citep{hui2005study}. The infrastructure of light rail and the organisation of missions allow to rapidly connect with the closest station, distributing a highly uniform accessibility for all districts of the territory. On the contrary in Zhuhai, the village of Tangjia is old, even anterior to the rest of Zhuhai, and has developed without any particular articulation with transportation infrastructures. The location of the tramway, which just opened, completes the trajectory of the new railway line, with an objective of reorganizing the North of Zhuhai, and in particular the High-tech Zone which extends from the North railway station (Zhuhai Bei) to Tangjia. Currently, the urban organisation is strongly imprinted with this unsynchronized development, since public transportation accessibility is still relatively low, bus lines being subject to an increasing congestion due to the strong increase in the number of cars. Furthermore, the exploitation of the tramway has been difficult, since the technology used with a third rail in the ground has been imported from Europe and had never been tested in such humidity conditions (personal communication with \noun{Yinghao Li}, July 2017), what lead to a questioning of the network plan in its entirety.

This fieldwork observations thus shows us that (i) under the same designation very different processes exist, and are extremely dependant to geographical, political and economical particularities; and that (ii) the development of a territory which is functional in terms of accessibility necessitates a fine articulation which seems to be the outcome of an integrated planning approach on the long time. In practice, such refined qualitative studies are necessary to complement the territorial diagnosis obtained with quantitative accessibility measures.

\section{Discussion}

\subsection{Long-term impacts of accessibility mutation}

The issue of long-term impacts of the evolution of accessibility landscape is at the core of the open question of interaction between networks and territories \citep{raimbault2018caracterisation}. Empirically, there does not seem to a general consensus on systematic effects, illustrating the problem of potential structuring effects of transportation infrastructures \citep{bonnafous1974methodologies}. For example, in the case of France, \cite{raimbault2017identification} identifies causal relationships between future network accessibility anticipation and socio-economic dynamics, whereas \cite{raimbault2018modeling} finds no relation between rail network growth and population growth on the long time.

In our study here for example, taking the Hong-Kong-Zhuhai-Macao bridge, the medium and long term impacts of this infrastructure are indeed difficult to estimate. In terms of short-term impact, \cite{wu2012impact} find patterns similar to the ones we estimate, i.e. a significant benefit for Zhuhai (and Hong-Kong that we did not take into account), and also immediate effects of traffic modification and economic impacts due to the toll or the increase of tourism. They mostly postulate the position of Zhuhai-Macao as a new pivot in the region. Even if it can be directly verified in terms of centrality and accessibility, it is not evident that this new position will influence particularly the socio-economic trajectory of Zhuhai. An increased particular political accompaniment implying an increased collaboration between Hong-Kong, Zhuhai and Macao will be important~\citep{zhou2016medium}. Immediate economic effects are expected, as an increase of Zhuhai residents working in Hong-Kong (Zhuhai inhabitants are the only ones in the region to benefit of a special card allowing them to regularly visit the Special Administrative Areas (source: fieldwork on 06/11/2016 with \noun{Cinzia Losavio}), but cases showing the contrary, such as investments from Hong-Kong towards the West of the Delta, have no reason to be systematic: the first case extends the already existing dynamic with Macao, the second is mostly to be constructed.


\subsection{Possible trajectories of the coupled network-territory system}

Generally, coupled trajectories for networks and territories can for the reasons above only be hypothesized. The use of simulation models is then an option to explore possible scenarios. A direction of exploration through modeling consists in considering the problem differently and to try to understand the dynamics of the metropolitan system in an integrated way, i.e. as a territorial system in our sense, in which the strong coupling between territory and network is operated through a proper ontology for governance entities. This approach was for example developed as a preliminary study by \cite{le2015modeling}.

At this stage, several assumptions for the future trajectory can be proposed: (i) in the worst case, networks do not respond well to the actual demand patterns, and territories do not particularly benefit from accessibility gains (this does not seem however to be the case as show the high use patterns of HSR and metro networks for example); (ii) accessibility gains are beneficial to territorial activities, but do not induce particular territorial bifurcations; (iii) the voluntary planning policy are actually effective, and significant mutations for some districts and cities are observed, together with significant modal shifts, enhancing an environmental and economic improvement. Given that such network developments in such a short time are without precedent, impacts may be significant and the last scenario seems to be plausible.

\subsection{Transferability of models}

We have studied here different cases of territorial development and infrastructure projects. The possibility of transfer of urban models (such as TOD), in the sense of the applicability of generic frameworks to different geographical contexts, is generally difficult. The synthesis of empirical conclusions obtained from very diverse case studies is also difficult. The East-Asian particularity has already been shown for the economic structure, and how it can not be interpreted in a simple way by a separation of microscopic and macroscopic processes as some quick and ideologically oriented readings may have done, such as the approach of the World Bank~\citep{amsden1994isn}. The comparability of urban systems is an open question at the core of issues for the Evolutive Urban Theory~\citep{pumain2015multilevel}. It is linked to the ergodic character of these systems: the ergodicity assumption postulates that the trajectory of a city in time captures the set of possible urban states, and also that different cities are different manifestations of the same stochastic process at different periods. In that case, an ensemble of cities would allow to understand their temporal trajectories. It is intuitively not the case, and urban systems would rather be non-ergodic~\cite{pumain2012urban}. Thus we will have to remain cautious for the generalization of conclusions.

\section*{Conclusion}

We have illustrated in this chapter how the deep mutations of transportation infrastructure networks can modify the accessibility landscapes, in the case of recent Chinese dynamics at different scales. We show that geographical balancing of accessibility is achieved and that highly accessible territories are produced. These mutations can be observed both from a quantitative and qualitative viewpoints, the second giving naturally more contrasted and specific conclusions. This suggest however models for sustainable transitions that most of territorial systems across the world still need to achieve.

\section*{Acknowledgements}


This research work was conducted in the context of the MEDIUM project (New pathways for sustainable development in Chinese MEDIUM size-cities). The author would like to acknowledge the support of the French National Center for Scientific Research (CNRS) and the UMR 8504 G{\'e}ographie-cit{\'e}s and to thank MEDIUM partners, in particular Sun-Yat Sen University. The MEDIUM project has received funding from the European Union under the External actions of the EU - Grant Contract ICI+/2014/348-005. The author would like to thank personally Yinghao Li (LEESU, ENPC) and Cinzia Losavio (G{\'e}ographie-cit{\'e}s, CNRS) for fieldwork insights and guidance.



\end{document}